# Vertically-Oriented Graphene Oxide Membranes for High-Performance Osmotic Energy Conversion

Zhenkun Zhang[1], Wenhao Shen[2], Lingxin Lin[1], Mao Wang[2], Ning Li[1], Zhifeng Zheng[1], Feng Liu[2,3]*, and Liuxuan Cao[1]*

1. College of Energy, Xiamen University, Xiamen, Fujian 361005, P. R. China.
2. State Key Laboratory of Nuclear Physics and Technology, Peking University, 100871 Beijing, P. R. China.
3. Center for Quantitative Biology, Peking University, 100871 Beijing, P. R. China.
E-mail: caoliuxuan@xmu.edu.cn,
liufeng-phy@pku.edu.cn

Keywords: Vertically-Oriented, Graphene oxide membrane, Ion selectivity, High ion permeability, Osmotic energy conversion

**Abstract**

Reverse electrodialysis is a promising method to harvest the osmotic energy stored between seawater and freshwater, but it has been a long-standing challenge to fabricate permselective membranes with the power density surpassing the industry benchmark of 5.0 W m$^{-2}$ for half a century. Herein, we report a vertically-oriented graphene oxide membranes (V-GOMs) with the combination of high ion selectivity and ultrafast ion permeation, whose permeation is three orders of magnitude higher than the extensively studied horizontally stacked GOMs (H-GOMs). By mixing artificial seawater and river water, we obtained an unprecedented high output power density of 10.6 W m$^{-2}$, outperforming all existing materials. Molecular dynamics (MD) simulations reveal the mechanism of the ultrafast transport in V-GOMs results from the quick entering of ions and the large accessible area as well as the apparent short diffusion paths in V-GOMs. These results will facilitate the practical application of osmotic energy and bring innovative design strategy for various systems involving ultrafast transport, such as filtration and catalysis.

**Introduction**

To address the challenge of global warming and environmental deterioration, renewable energies have become an urgent demand for the sustainable development of human society[1,2]. Osmotic energy stored as the form of salinity difference between seawater and freshwater, is a completely clean energy source without any pollution and carbon dioxides emission[3-5]. Reverse electrodialysis, a conventional method used to retrieve salinity gradient energy, is constrained by the poor performance of permselective membranes during the past half a century[6]. Inspired by the biological ion channels, artificial nanopores are used for osmotic energy conversion owing to the exceptional ion transport properties on nanoscale[7]. The nanofluidic reverse electrodialysis system (NREDS) was realized on many kinds of single pore model[8-10], indicating great potential superior to conventional commercial materials. Towards practical application, the NREDS were widely studied in a variety of porous materials, including polymeric membranes[11,12], inorganic carbon materials[13,14], silicon-based materials[15], aluminum oxide (AAO) template[16], compound materials[17-19] and stacked 2D materials[20]. Up to now, the highest record by mixing seawater and river water is 4.1 W m$^{-2}$ obtained in MXene/Kevlar nanofiber composite membranes[21]. Although it largely outperforms commercial ion exchange membranes by nearly an order of magnitude, the power density is still below the commercialization benchmark of 5.0 W m$^{-2}$ in the seawater/fresh water systems[22,23].

From the aspects of fundamental transport mechanism, the essential approach to promote the performance of nanoporous membrane is to break through the tradeoff between the ion permeability and selectivity[24-26]. The recent rise of 2D materials provide an encouraging solution as the fast ion transport was observed in the interstitial space between restacked 2D nanosheets combined with high ion selectivity[27]. More interestingly, further researches reveal the unidirectional interlayer paths of vertically-oriented 2D nano-sheet structure enable an ultrafast migration of ions extremely superior to the counterpart with horizontally stacked structure. This exceptional character has been employed to fabricate high-performance supercapacitors and electrodes with rapid electrolyte ion diffusion, high areal capacitance and fast response[28-30]. It is expected that the 2D membrane comprised by highly vertically-oriented flakes should have outstanding ion permeability performance. However, the vertically-oriented 2D membrane is still absent but urgently demanded from either mechanism research or practical application.

Herein, we report a vertically-oriented graphene oxide membranes (V-GOMs) prepared through vacuum filtration combined with MEMS fabrication technology. Benefit from the vertically-oriented unidirectional lamellas, the ion has ultrafast ion permeation in V-GOMs, with three orders of magnitude higher than that in H-GOMs. Due to the extraordinary combination of selectivity and permeability, the V-GOMs can achieve an unprecedented output power density of 10.6 W m$^{-2}$ by mixing seawater and river water, which largely outperforms the existing materials and beyond the critical value of 5.0 W m$^{-2}$ for industrial development requirement. The theoretical analysis and molecular dynamics simulations reveal the ultrafast ion transport mechanism in V-GOMs. Besides the apparent short diffusion paths in unidirectional lamellas, the other two factors, the loading time of ions entering GOMs without moving back and the accessible area on the membrane surface, paly more essential roles in the ultrafast ion transport in V-GOMs. These findings can greatly promote the practical applications of osmotic energy and open an innovative avenue towards various systems that involve ultrafast transport, such as filtration and catalysis.

**Results**

**Fabrication and characterization of GOMs.** Fig. 1a shows the fabrication processes for V-GOMs. The GOMs were formed through vacuum filtration. They were further cut into proper pieces and encapsulated by epoxy glue. The method of mechanical dicing, polishing and ion thinning were used to make the vertically-oriented GO structures exposed from the cured epoxy glue and to reduce the membrane thickness. In this way, we can obtain the V-GOMs with appropriate thickness and uniform lamellar microstructures (Fig. 1b). The Atomic Force

Microscope (AFM) observation suggests that the lateral size distribution of these GO sheets ranges from 400 nm to 1000 nm (Fig. 1c and Supplementary Fig. 1) and the average height is 0.9 nm (Fig. 1d). X-ray diffraction (XRD) patterns indicate that the interlayer spacing of the GOMs is 0.86 nm (Fig. 1e). Fourier transform infrared spectroscopy (FT-IR) shows the presence of multiple oxygen-containing functional groups in graphene oxide (Fig. 1f). These oxygen-containing functional groups can be quantitatively analyzed by X-ray photoelectron spectroscopy (XPS)[31], in which the carboxyl acid group is 1.2% of the total carbon content (Supplementary Fig. 3). The surface charge density of GO sheet is -73.8 mC m$^{-2}$ after the complete ionization of carboxyl acid group, which is in agreement with the previous results[20]. The GOMs are hydrophilic with surface contact angles of 55.6° (Fig. 1g). The zeta potential of the GO aqueous solution pulverized by ultrasonication shows the stability over the pH range of 3–11 (Fig. 1h)[32]. These characterization results are consistent with previous literature reports[20,31-33].

**Ultrahigh output power density of V-GOMs.** We measured the output power of V-GOMs to harvest the osmotic energy stored in the artificial seawater and river water (Fig. 2a). The V-GOMs were placed between 500 mM NaCl and 10 mM NaCl, which is common used in the previous researches to simulate the seawater and river water[5,34]. Electric power could be generated from the osmotic energy because of the charge separation in the ion-selective nanochannels[35]. The membrane potential of 79.1 mV and diffusion current density of 570 A m$^{-2}$ were respectively read from the intercepts on the horizontal and longitudinal coordinate (Fig. 2b). Reference electrodes were applied in electrical measurement to eliminate the contribution of redox potential generated by the unequal potential change at the electrode–solution interface[10,21].

The harvested electric power can be output to an external circuit (Fig. 2c). Through measuring the electrical signals in the electronic load, the electric power ($P_R$) consumed on the external resistance ($R$) can be directly calculated by $P_R = I^2 \times R$. The current decreases gradually with the increment of external resistance (Supplementary Fig. 4). And the output power achieves its peak value when the load resistance is equal to the internal resistance of membrane. For V-GOMs with the thickness of 358 µm, when the widths (*W*) of V-GOMs are 1.37 µm, 3.11 µm and 8.82 µm, the output powers reach the maximum when the external resistance is about 100 kΩ, 45 kΩ and 15 kΩ, respectively.

The obtained output power density of V-GOMs can achieve an unprecedented high value of 10.6 W m$^{-2}$, which is the highest reported record among all existing materials used to harvest osmotic energy, including PCTE[36], UFSCNM[17], PSS/MOF[18], IDM[13], PPy[12], PET-BCP[19], MKNCM[21], H-GOMs[20], CMI[13] and FKS[13] (Fig. 2d). Most importantly, this extremely high output power density apparently exceeds the commercialization benchmark of 5.0 W m$^{-2}$ for the first time, which probably promotes the industrialized utilization of salt difference energy[5]. To fairly reflect the properties of materials, the data presented in Fig. 2d are all under the same concentration difference of 500 mM | 10 mM NaCl, close to the concentration condition of most river water and seawater[20,21].

**Enhancement of the power density of V-GOM.** The power generation performance of V-GOMs can be further improved through the increment of concentration difference and pH value of electrolyte solution, including the diffusion current density (Fig. 3a) and membrane potential (Supplementary Fig. 5), which is consistent with the conventional viewpoint[37]. Notably, when the concentration difference is increased by 20 times to 1000 mM|1 mM NaCl and pH is increased to 11 at room temperature, the output power density of V-GOMs can be enhanced by nearly 3 times to 29.1 W m$^{-2}$ (Supplementary Fig. 6). Moreover, if the Na$^+$ ions are replaced by K$^+$ ions with higher mobility, the power density could be further enhanced to 34.3 W m$^{-2}$. This value surpasses the recent highest record set by 3 µm thick MXene lamellar showing a power density of 21 W m$^{-2}$ under the same condition (1000 mM|1 mM KCl at pH=11)[38].

Besides, the output power density of V-GOMs can also be significantly enhanced through reducing the membrane thickness. As shown in Fig. 3b, the power generation shows classical Ohm-type membrane-thickness dependence: the diffusion current density linearly decreases with the reduction of membrane thickness from 1800 μm to 350 μm; the membrane potential keep constant at about 80 mV with the change of membrane thickness. This is easy to understand because the membrane thickness only affects transmembrane resistance instead of the charge selectivity[39]. Accordingly, the total power density grows lineally with the thinning of V-GOMs (Supplementary Fig. 7). Of note, the smallest thickness of V-GOMs in this work is 358 μm owing to the limitation of our current preparation technology. Even in such thick membrane compared to the existing materials[17,21,40], the output power density obtained in 358 μm-thick V-GOMs is higher than 10 W m$^{-2}$ in the standard test condition (500 mM | 10 mM NaCl at pH=6). If the membrane thickness were reduced to below 30 μm, the power generation of V-GOMs has nearly ten times growth potential since the previous research has pointed out that the power density could be linearly enhanced within this thickness range[39].

**Scalability and aqueous stability.** The total output power of V-GOMs can be easily promoted by enlarging the effective membrane area. As shown in Fig. 3c, the diffusion current driven is facilitated by the increasing lengths ($L$) from 1.01 mm to 4.46 mm. As expected, the V-GOMs provide stable membrane potential of about 80 mV because of the constant ion selectivity. Similarly, the prolonged width ($W$) also raises the diffusion current linearly (Supplementary Fig. 9). In fact, the output electric powers show an excellent linear relationship with respect to the testing area of V-GOMs while the power density keeps constant (Table 1).

The V-GOMs also possess the excellent stability in aqueous solution environment as well as the high power density. The measure of electrical power generation lasted for more than 100 hours. The initial membrane potential is 83 mV, and the magnitude of voltage drop can be controlled within 10% during the 100-hour test. Meanwhile, the maximum output power density is basically maintained above 10.5 W m$^{-2}$ (Fig. 3d), suggesting the satisfying stability in practical application.

**High ionic permeability and selectivity.** To understand the origin of the high performance of V-GOMs in osmotic energy conversion, we systematically investigate ion selectivity and permeability, the two most important factors affecting the output power, in V-GOMs with the extensively used H-GOMs[31,41,42] as a comparison. We prepared V-GOMs and H-GOMs with the same fabrication condition to ensure they have the identical interlayer spacing and chemical composition, which accounts for the similar ion selectivity in V-GOMs and H-GOMs[43] (Supplementary Fig. 14). However, the different orientations of 2D flakes induce two types of ion transport modes across the membranes (Fig. 4a). As shown in Fig. 4b, the ionic conductance declines with the electrolyte concentration reducing from 1 M to 0.1 mM, suggesting the electric-field-induced ion transportation. When the ion concentration further decreases to lower than 0.1 mM, the ionic conductance reaches a plateau, indicating the strong surface-charge-governed ion transporting behavior, which is consistent with those reported in the literatures[20]. Intriguingly, the ionic conductance in V-GOMs is several hundred times higher than that in H-GOMs in all testing concentration conditions, which is consistent with the data plotted in Fig. 2d. The effective testing areas of H-GOMs and V-GOMs are 200 μm×200 μm and 1.14 mm×3.11 μm, respectively. The data plotted in Fig. 4b is the result of normalization of testing area. Besides, the diffusion current density in V-GOMs driven by the salinity gradient also extremely surpass those in H-GOMs, which is nearly 5000 times enhanced in V-GOMs (Fig. 4c).

We further analyzed the ion selectivity of V-GOMs through measuring the IV curve in NaCl solutions with highly different concentrations (1 M |1 μM). Because of the imbalanced electrolyte solution in the two sides of the membrane, the major ionic carriers across the membrane come from the left reservoir. Thus, the positive and negative ionic currents are dominated by $Na^+$ and $Cl^-$, respectively. As shown in Fig. 4d, the tested $Na^+$ current is 10.9 times higher than the $Cl^-$ current. In this way, the cation transference number ($t_+$) of the V-GOMs can be calculated to be 0.916 by the formula of $t_+ = I_+/(|I_+| + |I_-|)$[44], which indicates the strong cation selectivity of V-GOMs.

**Ultrafast transport mechanism.** To reveal the origin of ultrafast ion permeation in V-GOMs, we performed MD simulations to explore the mechanism of ionic migration through V-GOMs and H-GOMs[45]. We constructed an atomic model of the GOM with the interlayer spacing of 1 nm and surface charge density and elemental composition similar to experimental samples (Fig. 5a). The MD simulations show that the transport rate of $Na^+$ ions through V-GOMs normalized by the surface area is 190 times larger than that of H-GOMs (Fig. 5b). It is easily thought to stem from the different passing paths inside V-GOMs and H-GOMs. During the ions passing through H-GOMs, they have to take zigzag trajectories to go through the gaps between adjacent layers (Supplementary Fig. 16a). In sharp contrast, most of the ions can pass straight through a single channel inside V-GOMs (Supplementary Fig. 16b). As a result, the ion average velocity along the permeation direction across H-GOMs remarkably slows down, which is more than 13.7 times less than that in V-GOMs (Fig. 5c).

Nevertheless, the differences in passing paths are not adequate to account for the disparities in permeability by more than two orders of magnitude. Another crucial factor responsible for the ultrahigh ionic permeation is identified as the rapid access provided by the unique geometric structure in V-GOMs. The access process of ions from the outer region into the channel can be evaluated as the loading time[46,47], which is the time elapse between two ions subsequently entering GOMs without moving out. The MD simulation illustrates that the loading time in H-GOMs is 34.8 times longer than that in V-GOMs (Fig. 5d).

There are two main factors contributing to this difference. First, $Na^+$ ions are more difficult to go inside H-GOMs than V-GOMs. During the entrance of $Na^+$ ions into H-GOMs, the tortuous geometric structures produce a strong barrier to impede the access of $Na^+$. For individual orifice, the relative ion numbers in H-GOMs decrease with the distance from the entrance much more drastically than that in V-GOMs (Fig. 5e). In the 0.8 nm depth from the orifices, there are nearly 33.8% ions staying inside V-GOMs. In contrast, only 0.2% ions remain in H-GOMs at the same depth. It results from the evident ion enrichment at the entrance of H-GOMs (Supplementary Fig. 17), which accordingly induces strong electrostatic repulsion and dehydration barrier for the Na$^+$ ions to enter. These huge disparities suggest the single channel in V-GOMs has higher transport efficiency rather than the counterpart in H-GOMs. Second, V-GOMs offer a much greater accessible area for ions to enter compared with H-GOMs. In the experiment, the proportion of entrance area to total surface area in V-GOMs can achieve to amazingly about 50%, which largely surpasses that in H-GOMs (< 1%). It is worthy to point out that, due to the limitation of computational capacity, the proportion of the entrance area of H-GOMs in the simulation is much larger than that in the experiment. Hence the transport ratio between V-GOMs and H-GOMs is underestimated in the simulation compared with the experiment.

**Discussion**

In summary, we fabricate V-GOMs through the vacuum filtration and MEMS fabrication technology. Owing to the extraordinary combination of ion selectivity and permeability, the V-GOMs can achieve an extremely high output power density of 10.6 W m$^{-2}$ from artificial seawater and river water, which apparently outperforms the existing materials and beyond the critical value of 5.0 W m$^{-2}$ for industrial development requirement.

This high power density was measured with the standard condition of artificial sea water and river water (500 mM|10 mM NaCl at pH=6), which is closest to the application environment of blue energy and therefore is adopted by most previous studies[5,8,10,13,20]. Actually, the output power density of V-GOMs can be further enhanced under the higher concentration difference and alkaline aqueous condition as well as adapting the electrolyte solution with higher diffusion coefficient of cations. For example, using 3000 mM|10 mM NaCl at pH=11, the output power density is 36.7 W m$^{-2}$, which is also the highest record of existing materials under the same testing conditions[38]. It is worth emphasizing that, the highest power density is realized in our V-GOMs with 358 μm in thickness, which is generally dozens of times higher than other high-performance materials.

The molecular dynamics simulations reveal the mechanism of ultrafast ion transport in V-GOMs (Fig. 5f). The penetrating passages in V-GOMs allow a short migration distance and efficiently promote the ionic transport rate. Furthermore, the unique geometric structures in V-GOMs lower the entrance barrier and provide the rapid and efficient access to the inside channel. These findings can bring innovative design strategy for various ion transport systems, including catalysis, chemical sensing, ion filtration, water purification and energy conversion.

**Methods**

**Fabrication of V-GOMs.** GO sheets stacked from bottom to top onto the supporting membrane by vacuum filtration. Afterwards, the GOMs were dried in air at room temperature to remove residual water. Then, the GOMs were cut into proper pieces and encapsulated by epoxy glue. The method of mechanical dicing, polishing and ion thinning were used to make the vertically-oriented GO structures exposed from the cured epoxy glue and to reduce the membrane thickness. In this way, we can obtain the V-GOMs with appropriate thickness and uniform lamellar microstructures. The detailed procedure is described in the Supplementary Information.

**Characterization**. The zeta potential and size distribution of GO colloids (0.1 mg mL$^{-1}$) were measured with Malvern Zetasizer NanoZS90. The size and thickness of GO sheets were characterized by atomic force microscope (FM-Nanoview 6800AFM). The interlayer distance was tested on a polycrystalline X-ray diffractometer with a Cu Kα radiation source (Rigaku Ultima IV). Fourier transform infrared spectrometer (Nicolet Is5) was employed to map the distribution of chemical bond on the surface of GO sheets. The hydrophilicity of GOMs was characterized by contact angle measuring instrument (JC200JC1). The microstructure was observed by the field emission scanning electron microscope (SUPRA 55 SAPPHIRE).

**MD simulations.** An atomic GOM model[48] was constructed based on the characterization parameters of the experimental samples (Supplementary Information). All MD simulations were performed using GROMACS4.6 with CHARMM36 force field and TIP3P water model. The simulation used Van der Waals interactions with a cutoff of 1 nm and Particle-Mesh Edward electrostatics. Periodic boundary conditions were applied to all three directions. The temperature was maintained at 300 K using v-rescale. Each ionic transport simulation was run for 40 ns with the constant number of particles, volume and temperature (NVT) ensemble after the solvation reaches equilibrium. Only later 35 ns of the simulation were used for data analysis to ensure that the ion transport reaches equilibrium.

**Acknowledgements**

This work was supported by the National Natural Science Foundation of China (11875076, 31670599, 11405143), the Fundamental Research Funds for the Central Universities of China (Grant No. 20720190127) and the Natural Science Foundation of Fujian Province of China (No. 2019J05015). The MD simulation was performed on the High Performance Computing Platform of the Center for Life Sciences, Peking University.

**Author contributions**

Z.Z. and W.S. contributed equally to this work. L.C. designed the experiments and analysis. Z.Z. and L.L. performed the experiment. N.L. and Z.Z. provides useful help for experimental testing. F.L. designed the theoretical simulation and analysis. W.S. and M.W. performed MD simulations. L.C. and F.L. prepared the manuscript. All authors discussed the results, commented on the manuscript and contributed to the writing of the paper.

**References**

1. Pacala, S. & Socolow, R. Stabilization wedges: Solving the climate problem for the next 50 years with current technologies. *Science* **305**, 968-972 (2004).
2. Lindley, D. The Energy Should Always Work Twice. *Nature* **458**, 138-141 (2009).
3. Logan, B. E. & Elimelech, M. Membrane-based processes for sustainable power generation using water. *Nature* **488**, 313-319 (2012).
4. Macha, M., Marion, S., Nandigana, V. V. R. & Radenovic, A. 2D materials as an emerging platform for nanopore-based power generation. *Nat. Rev. Mater.* **4**, 588-605 (2019).
5. Siria, A., Bocquet, M.-L. & Bocquet, L. New avenues for the large-scale harvesting of blue energy. *Nat. Rev. Chem.* **1**, 0091 (2017).
6. Pattle, R. Production of electric power by mixing fresh and salt water in the hydroelectric pile. *Nature* **174**, 660-660 (1954).
7. Guo, W., Tian, Y. & Jiang, L. Asymmetric Ion Transport through Ion-Channel-Mimetic Solid-State Nanopores. *Acc. Chem. Res.* **46**, 2834-2846 (2013).
8. Feng, J. et al. Single-layer $MoS_2$ nanopores as nanopower generators. *Nature* **536**, 197-200 (2016).
9. Siria, A. et al. Giant osmotic energy conversion measured in a single transmembrane boron nitride nanotube. *Nature* **494**, 455-458 (2013).
10. Guo, W. et al. Energy Harvesting with Single-Ion-Selective Nanopores: A Concentration-Gradient-Driven Nanofluidic Power Source. *Adv. Funct. Mater.* **20**, 1339-1344 (2010).
11. Ramirez, P. et al. Energy conversion from external fluctuating signals based on asymmetric nanopores. *Nano Energy* **16**, 375-382 (2015).
12. Yu, C. et al. A smart cyto-compatible asymmetric polypyrrole membrane for salinity power generation. *Nano Energy* **53**, 475-482 (2018).
13. Gao, J. et al. High-Performance Ionic Diode Membrane for Salinity Gradient Power Generation. *J. Am. Chem. Soc.* **136**, 12265-12272 (2014).
14. Lokesh, M., Youn, S. K. & Park, H. G. Osmotic Transport across Surface Functionalized Carbon Nanotube Membrane. *Nano lett.* **18**, 6679-6685 (2018).
15. Hwang, J., Kataoka, S., Endo, A. & Daiguji, H. Enhanced energy harvesting by concentration gradient-driven ion transport in SBA-15 mesoporous silica thin films. *Lab Chip* **16**, 3824-3832 (2016).
16. Kang, B. D., Kim, H. J., Lee, M. G. & Kim, D.-K. Numerical study on energy harvesting from concentration gradient by reverse electrodialysis in anodic alumina nanopores. *Energy* **86**, 525-538 (2015).
17. Xiao, K., Giusto, P., Wen, L., Jiang, L. & Antonietti, M. Nanofluidic Ion Transport and Energy Conversion through Ultrathin Free-Standing Polymeric Carbon Nitride Membranes. *Angew. Chem. Int. Ed.* **130**, 10280-10283 (2018).
18. Li, R., Jiang, J., Liu, Q., Xie, Z. & Zhai, J. Hybrid nanochannel membrane based on polymer/MOF for high-performance salinity gradient power generation. *Nano Energy* **53**, 643-649 (2018).
19. Zhang, Z. et al. Engineered Asymmetric Heterogeneous Membrane: A Concentration-Gradient-Driven Energy Harvesting Device. *J. Am. Chem. Soc.* **137**, 14765-14772 (2015).

20. Ji, J. et al. Osmotic Power Generation with Positively and Negatively Charged 2D Nanofluidic Membrane Pairs. *Adv. Funct. Mater.* **27**, 1603623 (2017).
21. Zhang, Z. et al. Mechanically strong MXene/Kevlar nanofiber composite membranes as high-performance nanofluidic osmotic power generators. *Nat. Commun.* **10**, 2920 (2019).
22. Sivertsen, E., Holt, T., Thelin, W. & Brekke, G. Pressure retarded osmosis efficiency for different hollow fibre membrane module flow configurations. *Desalination* **312**, 107-123 (2013).
23. Gerstandt, K., Peinemann, K. V., Skilhagen, S. E., Thorsen, T. & Holt, T. Membrane processes in energy supply for an osmotic power plant. *Desalination* **224**, 64-70 (2008).
24. Wang, L. et al. Fundamental transport mechanisms, fabrication and potential applications of nanoporous atomically thin membranes. *Nat. Nanotechnol.* **12**, 509-522 (2017).
25. Koros, W. J. & Zhang, C. Materials for next-generation molecularly selective synthetic membranes. *Nat. Mater.* **16**, 289-297 (2017).
26. Park, H. B., Kamcev, J., Robeson, L. M., Elimelech, M. & Freeman, B. D. Maximizing the right stuff: The trade-off between membrane permeability and selectivity. *Science* **356**, 1137 (2017).
27. Yang, J. et al. Photo-induced ultrafast active ion transport through graphene oxide membranes. *Nat. Commun.* **10**, 1171 (2019).
28. Seo, D. H., Han, Z. J., Kumar, S. & Ostrikov, K. Structure-Controlled, Vertical Graphene-Based, Binder-Free Electrodes from Plasma-Reformed Butter Enhance Supercapacitor Performance. *Adv. Energy Mater.* **3**, 1316-1323 (2013).
29. Xia, Y. et al. Thickness-independent capacitance of vertically aligned liquid-crystalline MXenes. *Nature* **557**, 409-412 (2018).
30. Zheng, S. et al. High Packing Density Unidirectional Arrays of Vertically Aligned Graphene with Enhanced Areal Capacitance for High-Power Micro-Supercapacitors. *ACS Nano* **11**, 4009-4016 (2017).
31. Chen, L. et al. Ion sieving in graphene oxide membranes via cationic control of interlayer spacing. *Nature* **550**, 380-383 (2017).
32. Baskoro, F. et al. Graphene oxide-cation interaction: Inter-layer spacing and zeta potential changes in response to various salt solutions. *J. Membr. Sci.* **554**, 253-263 (2018).
33. Voiry, D. et al. High-quality graphene via microwave reduction of solution-exfoliated graphene oxide. *Science* **353**, 1413-1416 (2016).
34. Zhu, X. et al. Unique ion rectification in hypersaline environment: A high-performance and sustainable power generator system. *Sci. Adv.* **4**, 1665 (2018).
35. Cao, L. et al. On the Origin of Ion Selectivity in Ultrathin Nanopores: Insights for Membrane-Scale Osmotic Energy Conversion. *Adv. Funct. Mater.* **28**, 1804189 (2018).
36. Kwon, K., Lee, S. J., Li, L., Han, C. & Kim, D. Energy harvesting system using reverse electrodialysis with nanoporous polycarbonate track-etch membranes. *Int. J. Energy Res.* **38**, 530-537 (2014).
37. Cao, L. et al. Towards understanding the nanofluidic reverse electrodialysis system: well matched charge selectivity and ionic composition. *Energy Environ. Sci.* **4**, 2259-2266 (2011).
38. Hong, S. et al. Two-Dimensional $Ti_3C_2T_x$ MXene Membranes as Nanofluidic Osmotic Power Generators. *ACS Nano* **13**, 8917-8925 (2019).

39. Cao, L. et al. Anomalous Channel-Length Dependence in Nanofluidic Osmotic Energy Conversion. *Adv. Funct. Mater.* **27**, 1604302 (2017).
40. Fu, Y., Guo, X., Wang, Y., Wang, X. & Xue, J. An atomically-thin graphene reverse electrodialysis system for efficient energy harvesting from salinity gradient. *Nano Energy* **57**, 783-790 (2019).
41. Joshi, R. K. et al. Precise and Ultrafast Molecular Sieving Through Graphene Oxide Membranes. *Science* **343**, 752-754 (2014).
42. Abraham, J. et al. Tunable sieving of ions using graphene oxide membranes. *Nat. Nanotechnol.* **12**, 546-550 (2017).
43. Vlassiouk, I., Smirnov, S. & Siwy, Z. Ionic selectivity of single nanochannels. *Nano Lett.* **8**, 1978-1985 (2008).
44. Cao, L., Guo, W., Wang, Y. & Jiang, L. Concentration-Gradient-Dependent Ion Current Rectification in Charged Conical Nanopores. *Langmuir* **28**, 2194-2199 (2012).
45. Dai, H., Xu, Z. & Yang, X. Water Permeation and Ion Rejection in Layer-by-Layer Stacked Graphene Oxide Nanochannels: A Molecular Dynamics Simulation. *J. Phys. Chem. C* **120**, 22585-22596 (2016).
46. Wang, P. et al. Ultrafast ion sieving using nanoporous polymeric membranes. *Nat. Commun.* **9**, 569 (2018).
47. Wen, Q. et al. Highly Selective Ionic Transport through Subnanometer Pores in Polymer Films. *Adv. Funct. Mater.* **26**, 5796-5803 (2016).
48. Willcox, J. A. L. & Kim, H. J. Molecular Dynamics Study of Water Flow across Multiple Layers of Pristine, Oxidized, and Mixed Regions of Graphene Oxide. *ACS Nano* **11**, 2187-2193 (2017).

**Figure 1 | Preparation and characterization of V-GOMs. a,** The fabrication process of V-GOMs. **b,** SEM image of V-GOMs showing compactly packed lamellar structures. **c, d,** A typical AFM observation of the GO sheets suggests that the thickness of individual GO sheets is about 0.9 nm. **e,** Sharp XRD peak at 10.3° indicates the uniform interlayer distance of 0.86 nm. **f,** FT-IR spectra indicates the chemical functional groups in synthetic GO sheets. **g,** The GOMs are hydrophilic with surface contact angles of 55.6°. **h,** Surface charge properties of GO colloids under varied pH conditions (0.1 mg mL$^{-1}$).

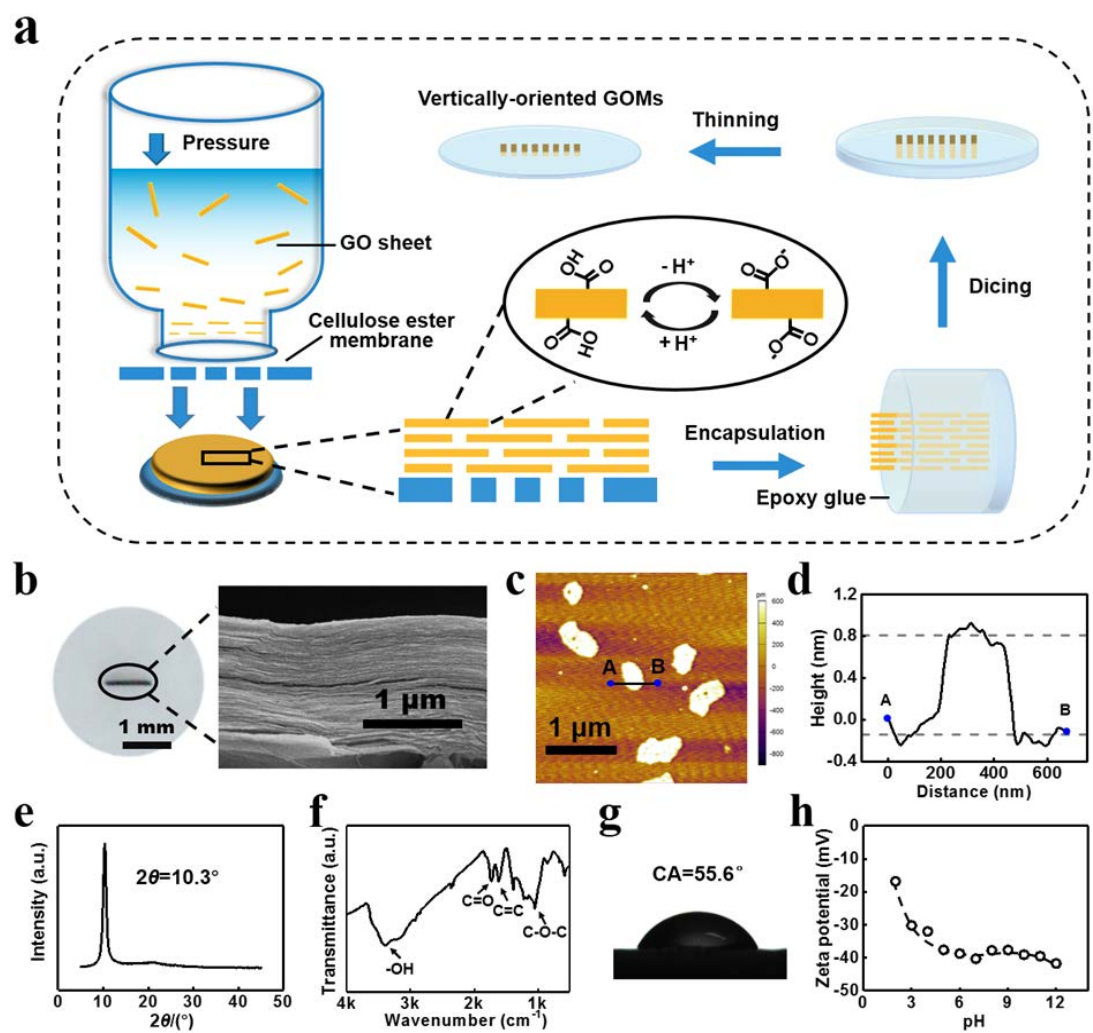

**Figure 2 | Ultrahigh output power density of V-GOMs. a,** Schematic of energy harvesting from the seawater and river water. **b,** IV curve of V-GOMs. **c,** The output power measured in external loads. The obtained output power density of V-GOMs is as high as 10.6 W m$^{-2}$. **d,** Benchmark of the V-GOMs with other reported osmotic energy conversion systems for the output power density and energy conversion efficiency. The V-GOMs have the highest output power density.

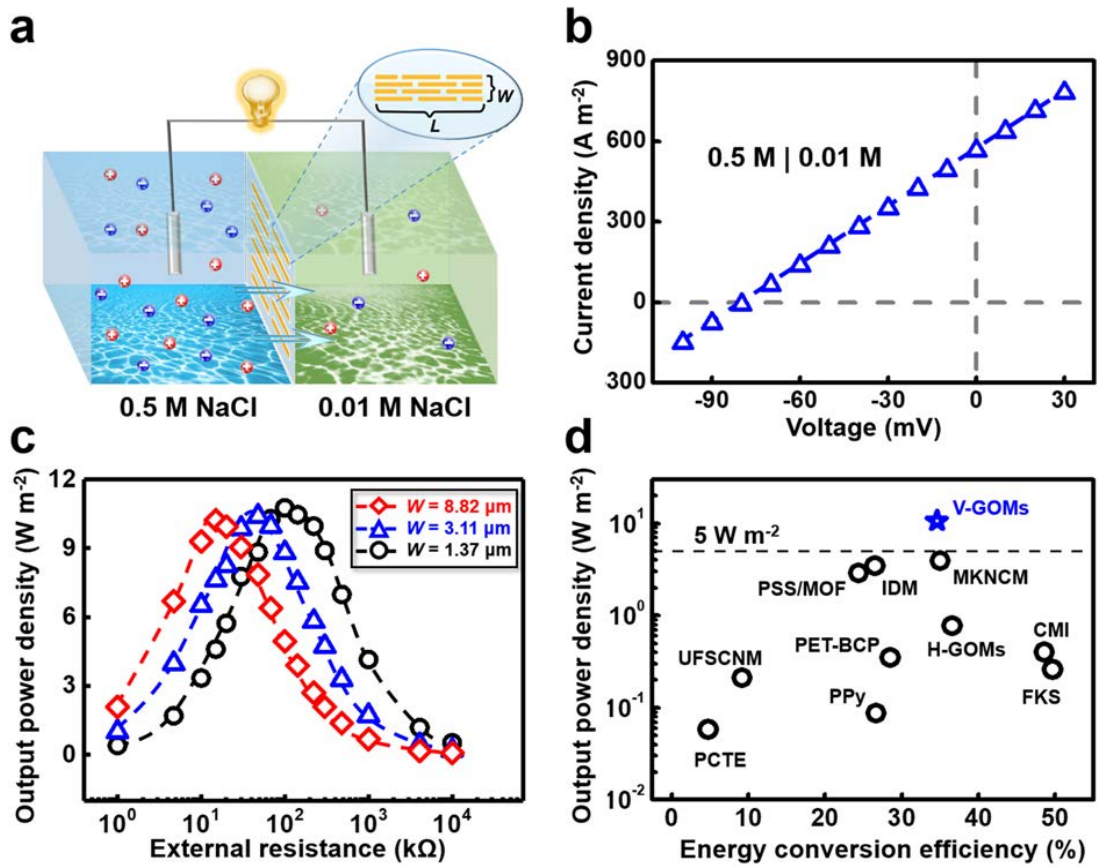

**Figure 3 | Energy conversion of V-GOMs in different conditions. a,** The diffusion current density increases with the applied concentration difference and pH value. **b,** The generated current density declines with the increasing membrane thickness, showing Ohm-like dependence. The membrane potential is not sensitive to the membrane thickness. **c,** The diffusion current increases with the membrane length (*L*), but the membrane potential remains stable. **d,** V-GOMs can keep stable performance in electrolyte solution for several days. The applied concentration difference is 0.5 M | 0.01 M NaCl.

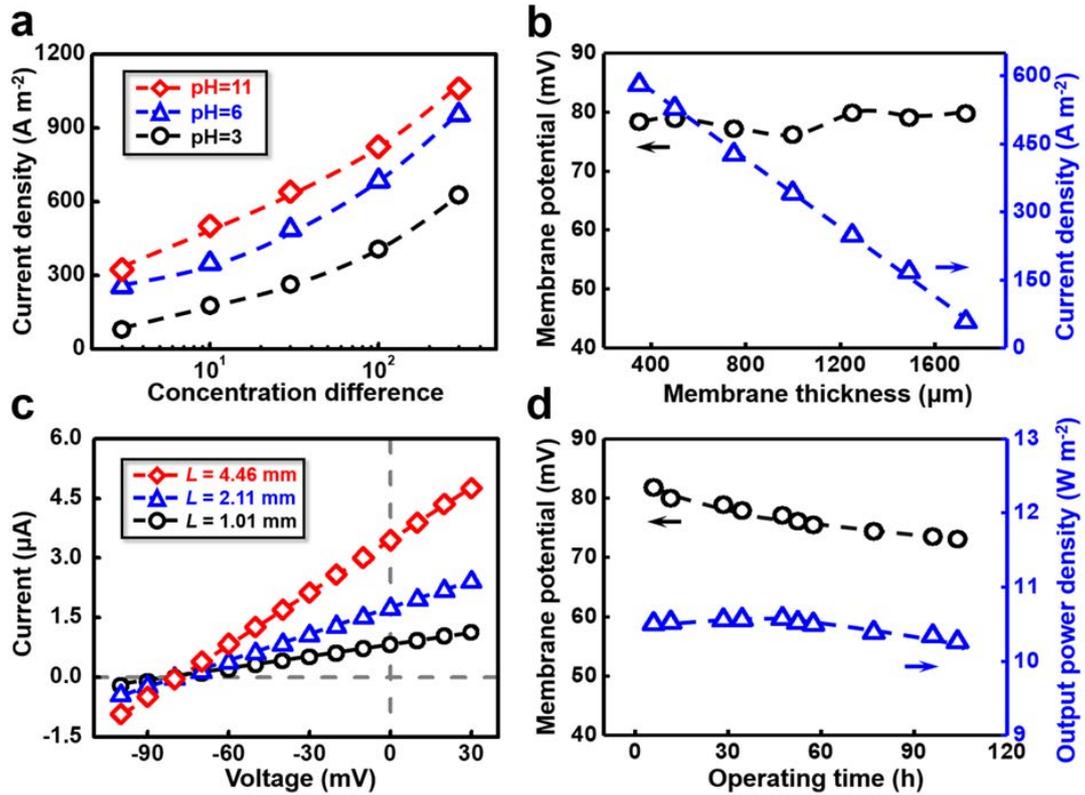

**Figure 4 | High permeability and selectivity of V-GOMs. a,** Schematic of ion transport in H-GOMs and V-GOMs. **b,** Ionic conductance of V-GOMs is several hundred higher than that of H-GOMs. The effective area of H-GOMs and V-GOMs is 200 × 200 μm and 1140 × 3.11 μm, respectively. **c,** Current density of V-GOMs driven by ion concentration gradient is more than that of H-GOMs with three orders of magnitude. The electrolyte solution is NaCl with low concentration side of 1 mM. **d,** Ion selectivity of the V-GOMs is verified by the ionic current under asymmetric concentration. The cation transference number ($t_+$) is 0.916.

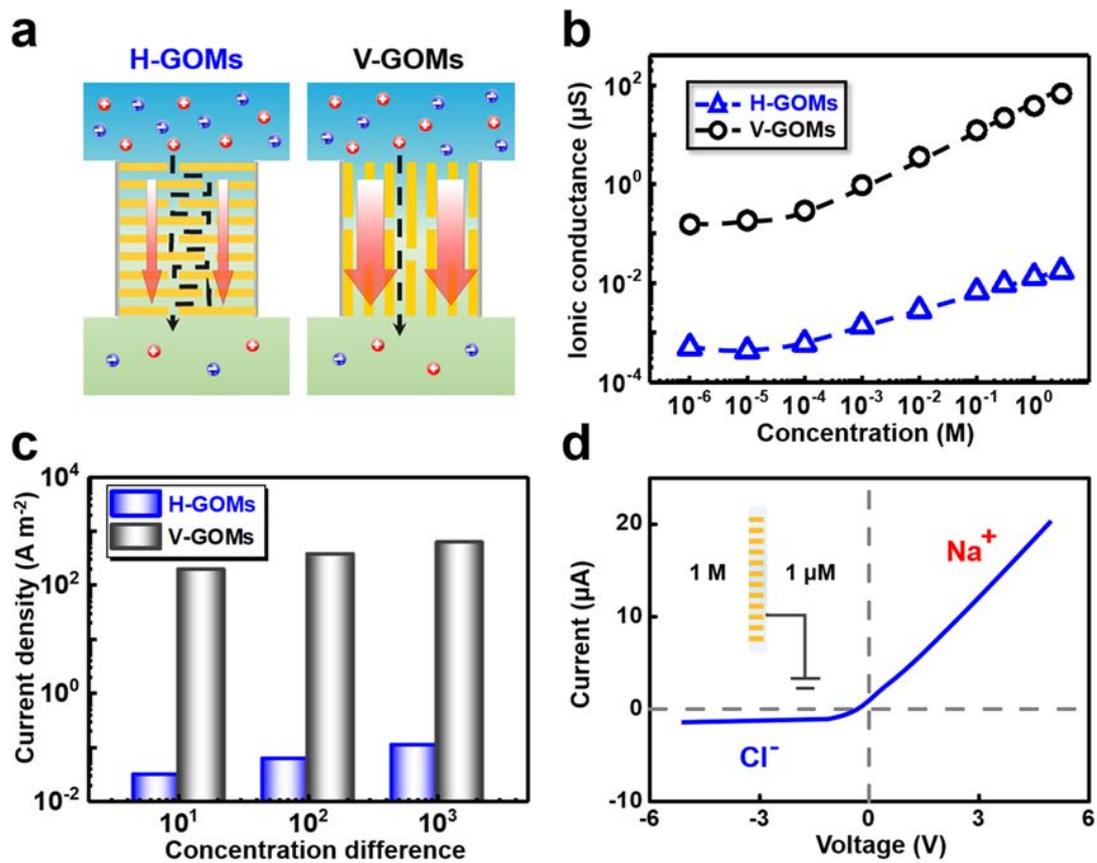

**Figure 5 | MD simulations of ionic transport through H-GOMs and V-GOMs. a,** The MD simulation model of the H-GOM (left) or V-GOM (right) is consisted of four GO flakes with the interlayer spacing of 1 nm. **b,** The transport rate of $Na^+$ ions through V-GOMs is 190 times larger than that of H-GOMs. **c,** The zigzag trajectories in H-GOMs remarkably slow down the ion average velocity, which is more than 13.7 times less than that in V-GOMs. **d,** The loading time in H-GOMs is more than 30 times longer than that in V-GOMs. **e,** For individual orifice, the relative ion number in H-GOMs decreases with the distance from the entrance much more drastically than that in V-GOMs. **f,** Schematic of the ultrafast ionic transport mechanism through V-GOMs superior to H-GOMs.

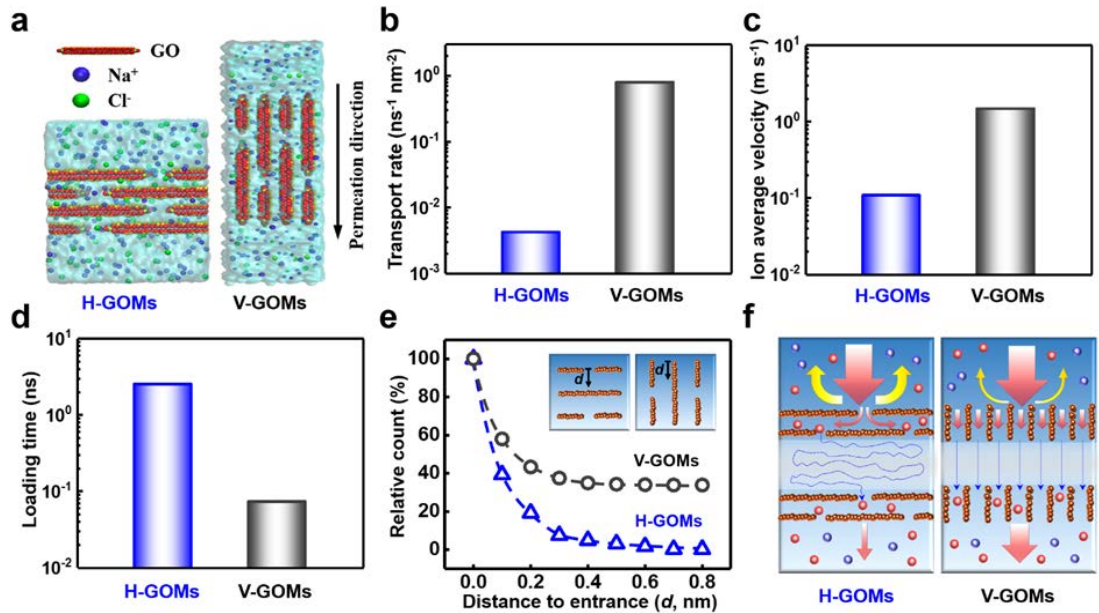

**Table 1.** The output power of different V-GOM samples. The applied solution condition is 0.5 M | 0.01 M NaCl. The membrane thickness is 358 μm.

| Sample | Length/(mm) | Width/(um) | Current/(μA) | Membrane potential/(mV) | Output power density/(W m$^{-2}$) |
|---|---|---|---|---|---|
| 1 | 1.01 | 1.37 | 0.816 | 79.6 | 10.8 |
| 2 | 2.11 | 1.37 | 1.72 | 78.6 | 10.6 |
| 3 | 4.46 | 1.37 | 3.45 | 79.1 | 10.2 |
| 4 | 1.06 | 0.80 | 0.505 | 80.2 | 10.9 |
| 5 | 0.98 | 1.37 | 0.804 | 79.3 | 10.7 |
| 6 | 0.96 | 3.11 | 1.76 | 78.4 | 10.4 |
| 7 | 1.04 | 6.08 | 3.45 | 80.4 | 10.1 |
| 8 | 1.02 | 8.82 | 5.23 | 77.8 | 10.3 |

# Supplementary Information

## Vertically-Oriented Graphene Oxide Membranes for High-Performance Osmotic Energy Conversion

Zhenkun Zhang[1], Wenhao Shen[2], Lingxin Lin[1], Mao Wang[2], Ning Li[1], Zhifeng Zheng[1], Feng Liu[2,3]*, and Liuxuan Cao[1]*

1. College of Energy, Xiamen University, Xiamen, Fujian 361005, P. R. China.
2. State Key Laboratory of Nuclear Physics and Technology, Peking University, 100871 Beijing, P. R. China.
3. Center for Quantitative Biology, Peking University, 100871 Beijing, P. R. China.

* Correspondence to: caoliuxuan@xmu.edu.cn,
liufeng-phy@pku.edu.cn

**This PDF file includes:**

Supplementary Figures 1-17

Supplementary Tables 1

Supplementary Texts 1-10

References

**Table of contents**

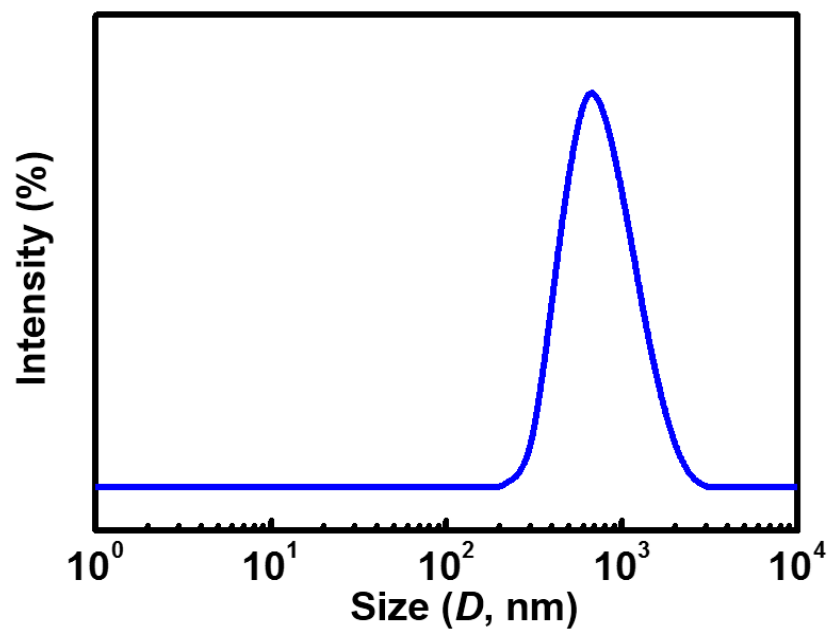

**Supplementary Fig. 1.** Size distribution of GO colloids was observed by a Malvern Zetasizer NanoZS90. The sizes of GO sheets ranges from 400 nm to 1000 nm. The concentration of GO dispersion is 0.1 mg mL$^{-1}$.

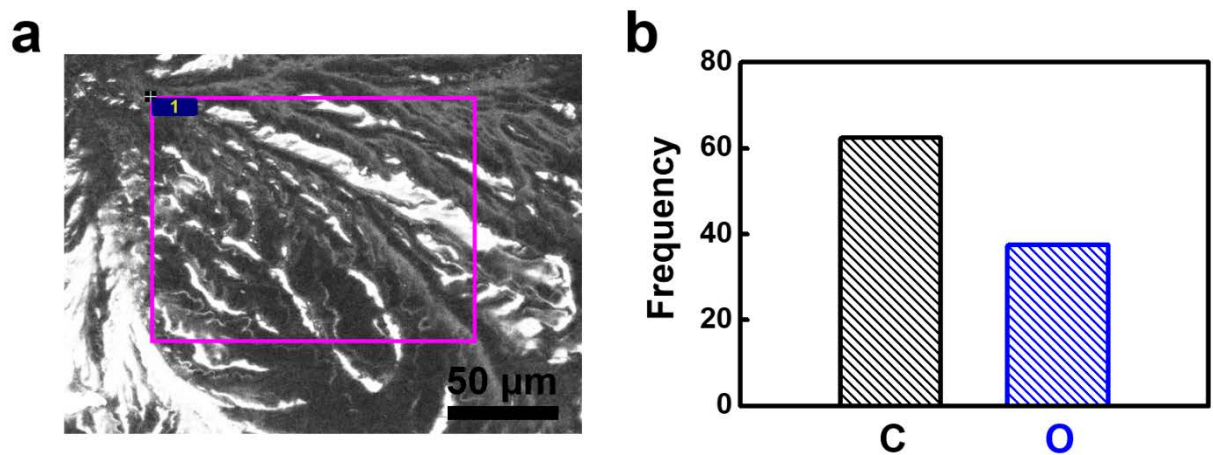

**Supplementary Fig. 2. a,** SEM observations on the surface of GOMs. **b,** The corresponding EDS analysis in the zone 1 from a. The graphite was partially oxidized to oxygen-containing groups.

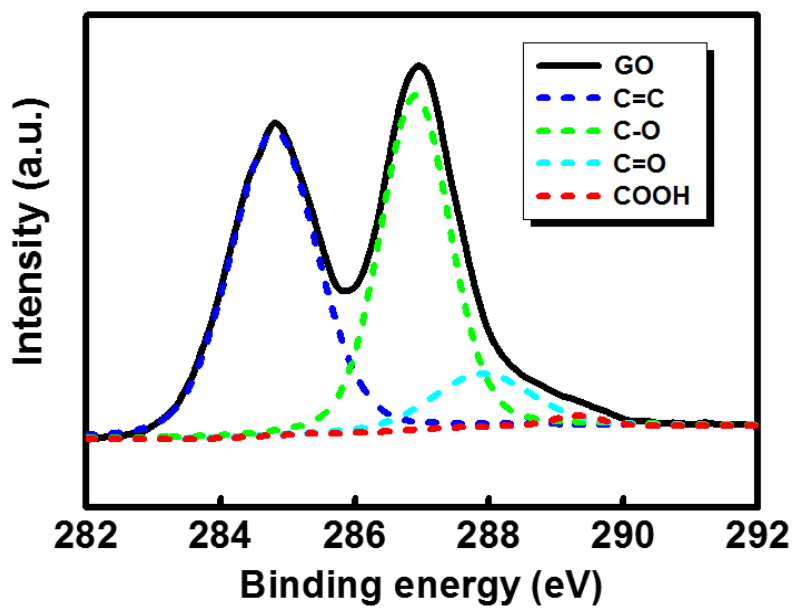

**Supplementary Fig. 3.** XPS analysis of GO. The composite peak of GO (solid line) was decomposed into four individual peaks (dashed lines) including C=C (284.8 eV), C-OH (286.9 eV), C=O (287.9 eV) and COOH (289.3 eV). The percentage of carboxylic acid group is 1.2% of the total carbon content.

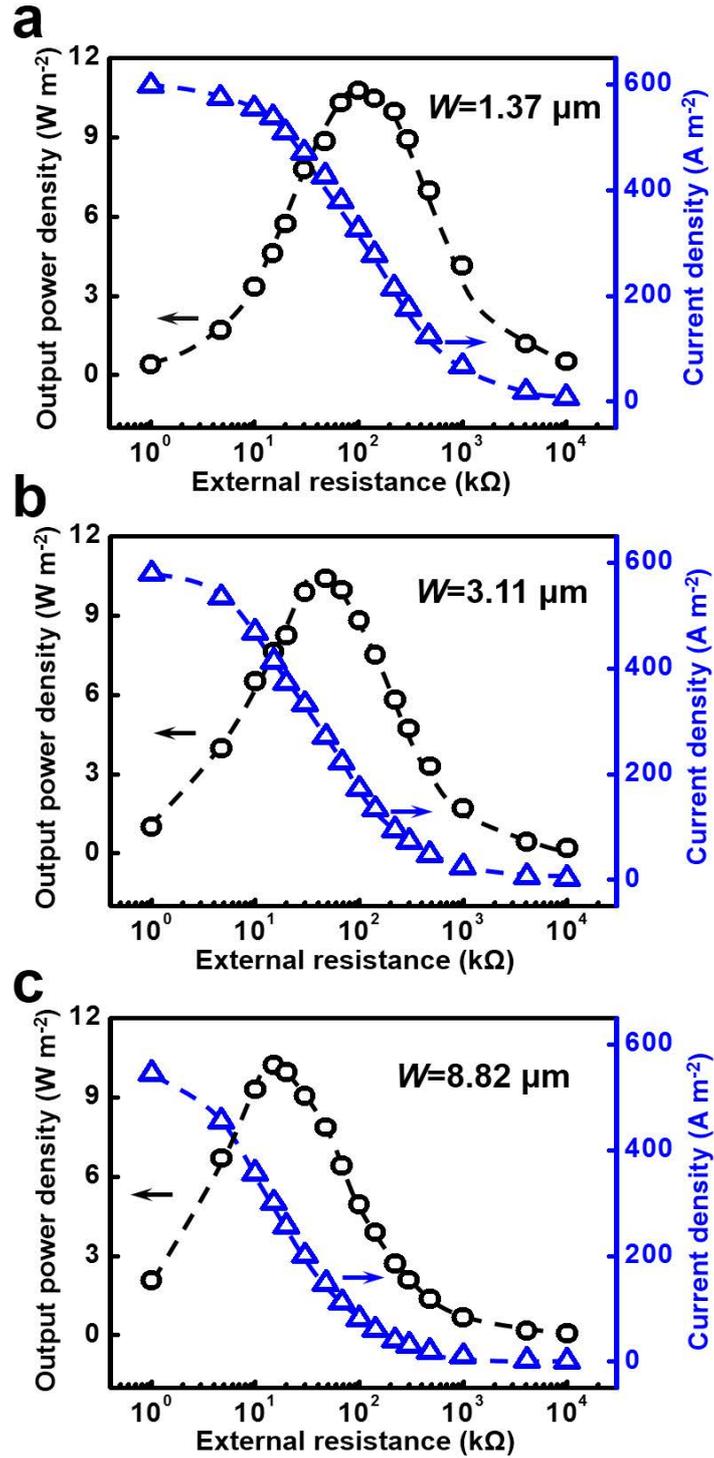

**Supplementary Fig. 4.** The output power densities and current densities measured in the external load when the areas of V-GOMs are **a,** 1.37 μm × 0.98 mm, **b,** 3.11 μm × 0.96 mm and **c,** 8.82 μm × 1.02 mm, respectively. Diluted (LC) and concentrated (HC) NaCl solution remains at 10 mM and 500 mM, respectively. These results correspond to Fig. 2c.

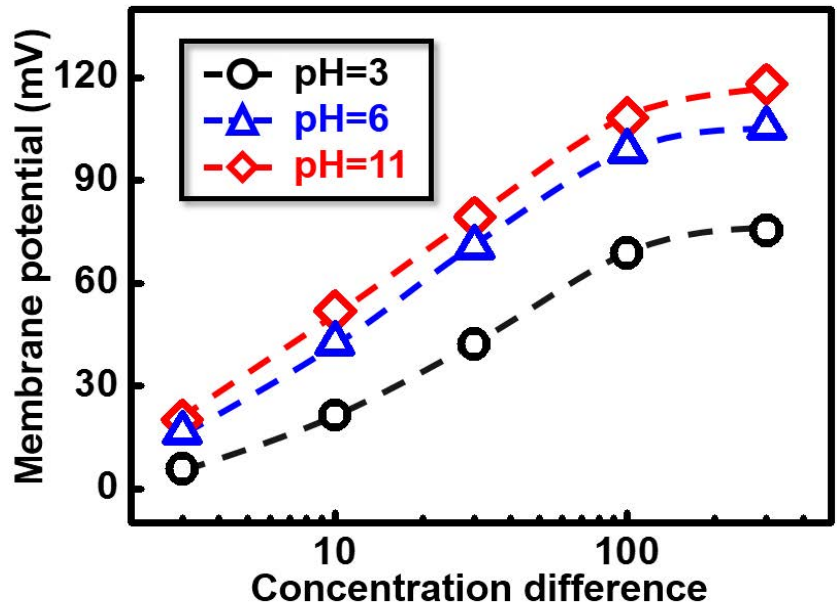

**Supplementary Fig. 5.** The membrane potential of V-GOMs increases with the pH value and ion concentration difference.

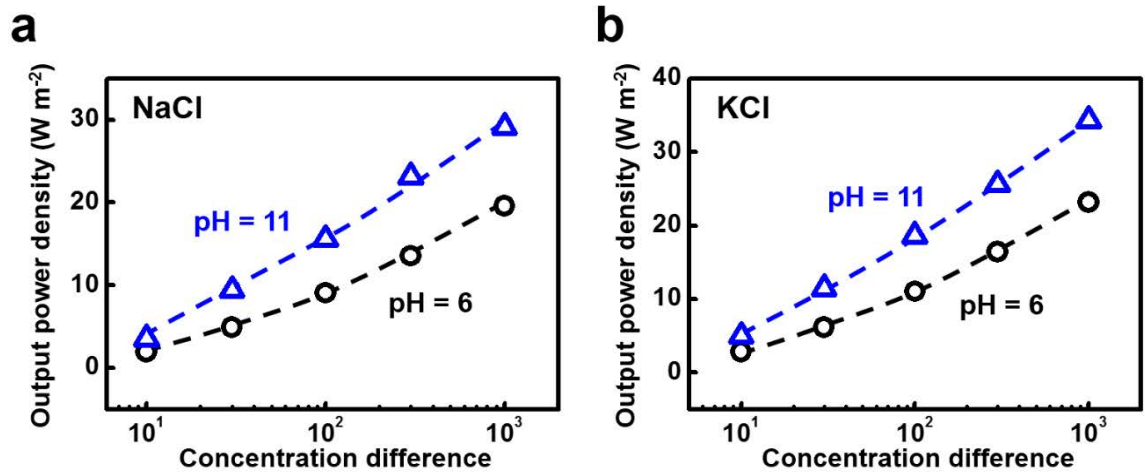

**Supplementary Fig. 6.** Output power density under different electrolyte conditions. (a) The output power density in NaCl solutions with different pH conditions. (b) The output power density in KCl solutions under different pH conditions.

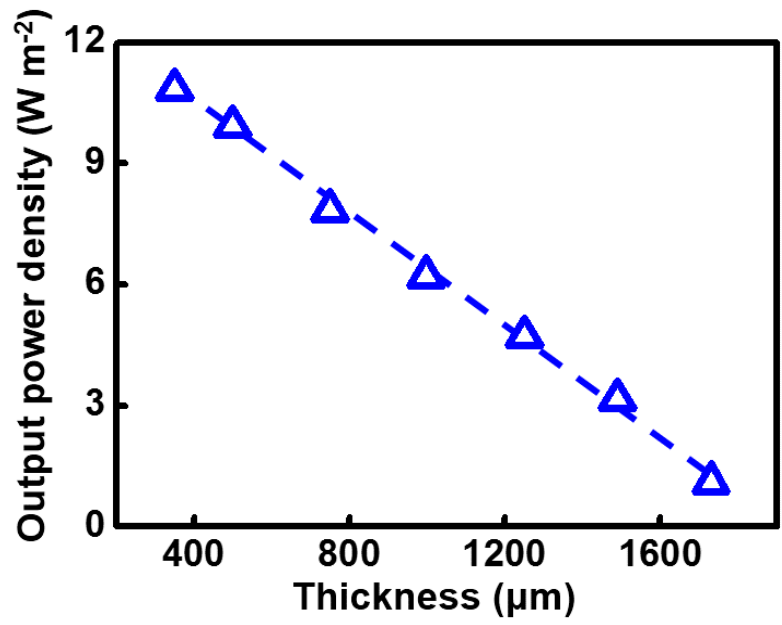

**Supplementary Fig. 7.** Under 50-fold concentration difference of NaCl solution, the output power density linearly decreases with the increment of the thickness of V-GOMs.

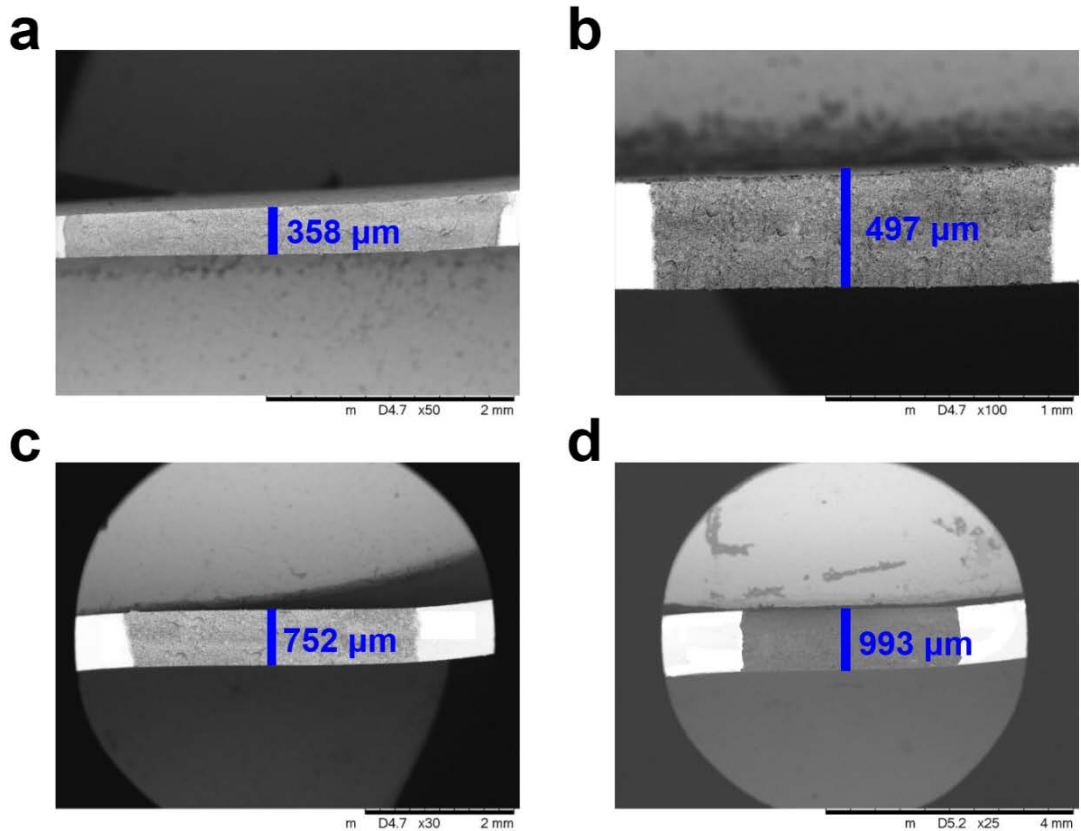

**Supplementary Fig. 8.** SEM observations of the profile of V-GOMs. The height of the profile corresponds to the thickness of V-GOMs. The thickness of V-GOMs is **a,** 358 μm, **b,** 497 μm, **c,** 752 μm and **d,** 993 μm, respectively.

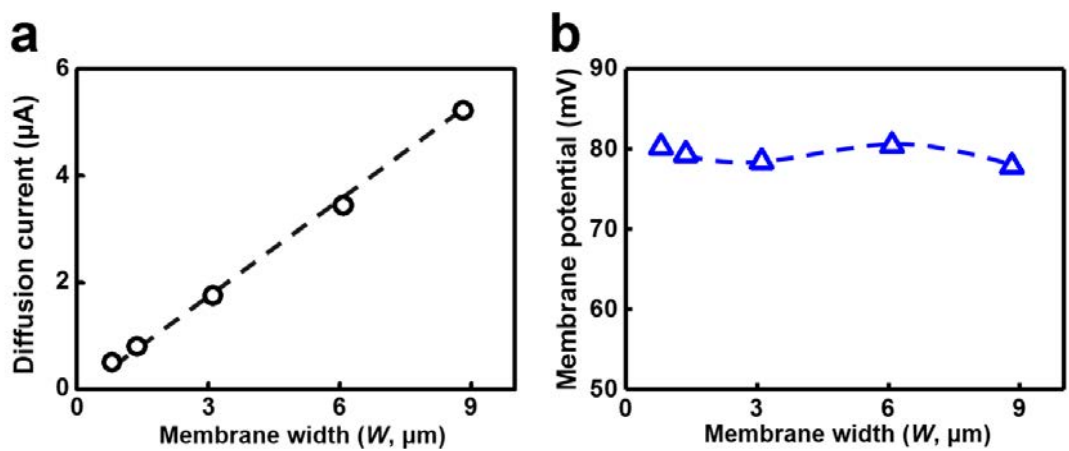

**Supplementary Fig. 9. a,** The prolonged width ($W$) raises the diffusion current linearly. **b,** The membrane potential basically maintains at 80 mV while the the widths of V-GOMs ranges from 0.5 μm to 9 μm. The applied solution is 50-fold NaCl solution.

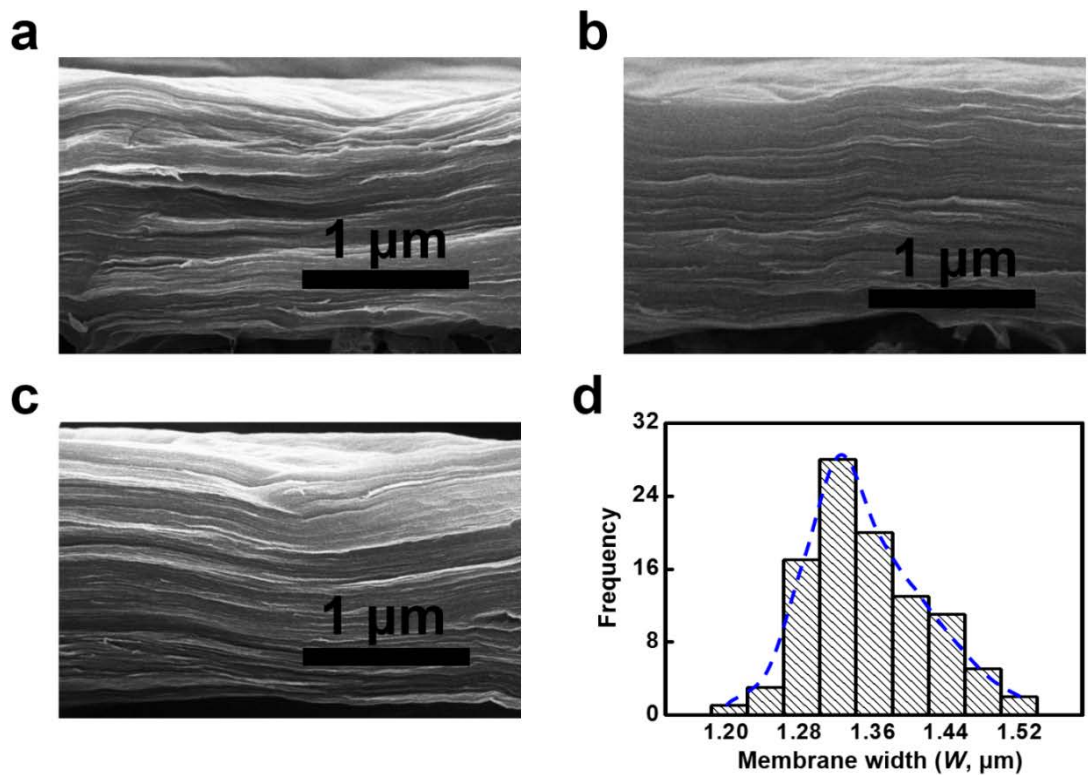

**Supplementary Fig. 10.** SEM observations to measure average width of the cross section of V-GOMs. **a, b, c,** The typical SEM images of the cross section in V-GOMs with the width (*W*) of 1.37 μm. **d,** Statistical results over 100 samples show that the width distribution of V-GOMs is 1.37 ± 0.17 μm.

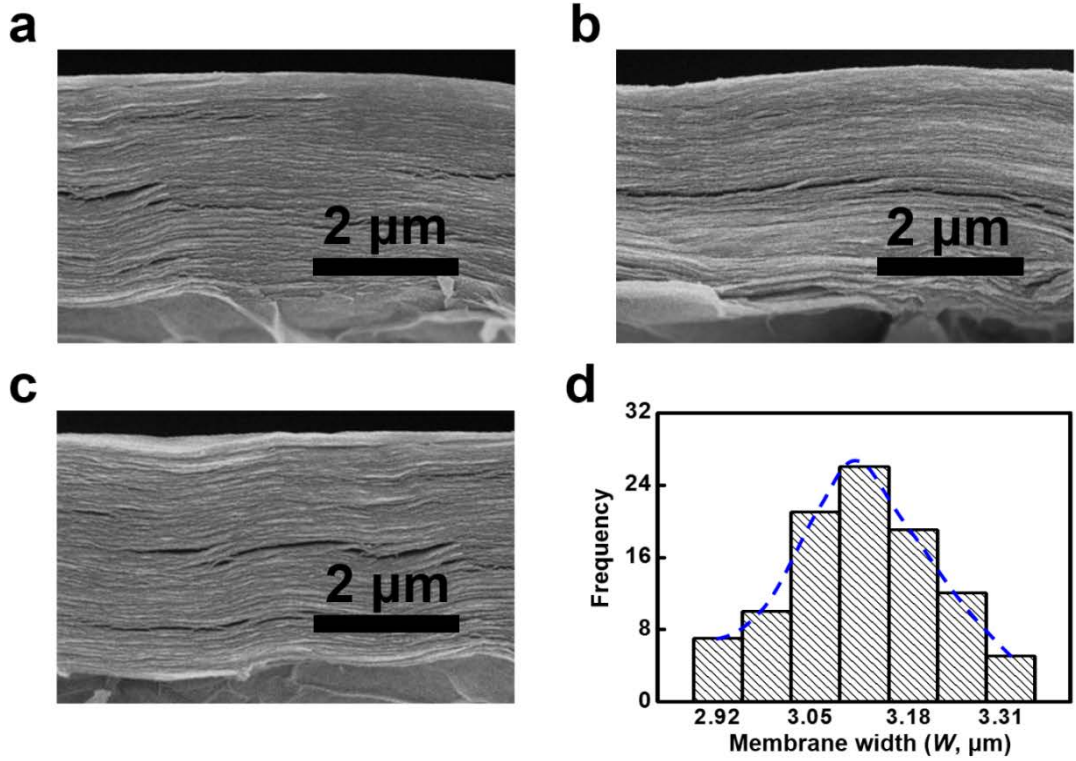

**Supplementary Fig. 11.** SEM observations to measure average width of the cross section of V-GOMs. **a, b, c,** The typical SEM images of the cross section in V-GOMs with the width (W) of 3.11 μm. **d,** Statistical results over 100 samples show that the the width distribution of V-GOM is 3.11 ± 0.19 μm.

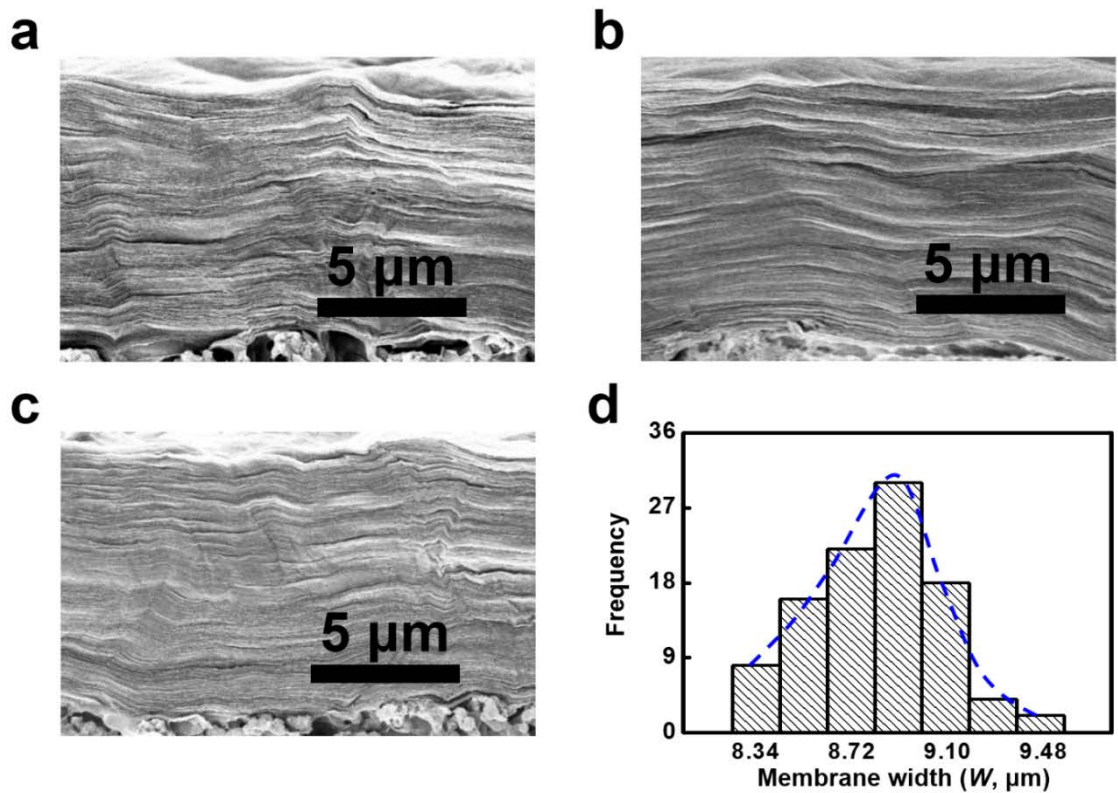

**Supplementary Fig. 12.** SEM observations to measure average width of the cross section of V-GOMs. **a, b, c,** The typical SEM images of the cross section in V-GOMs with the width (W) of 8.82 μm. **d,** Statistical results over 100 samples show that the the width distribution of V-GOMs is 8.82 ± 0.48 μm.

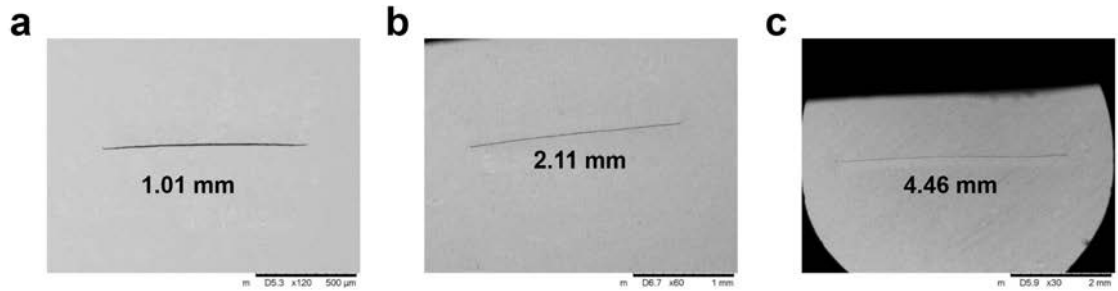

**Supplementary Fig. 13.** SEM observations to measure the average length (*L*) of the cross section of V-GOMs. **a, b, c,** The typical SEM images of the cross section in V-GOMs. The length of V-GOMs is 1.01 mm (a), 2.11 mm (b) and 4.46 mm (c), respectively.

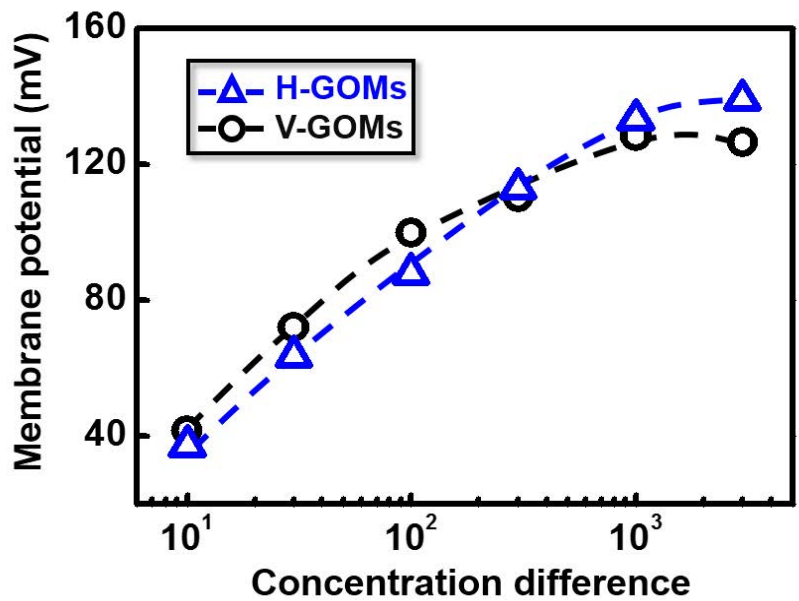

**Supplementary Fig. 14.** The membrane potentials of H-GOMs and V-GOMs are similar under the same concentration difference, indicating the same cation selectivity.

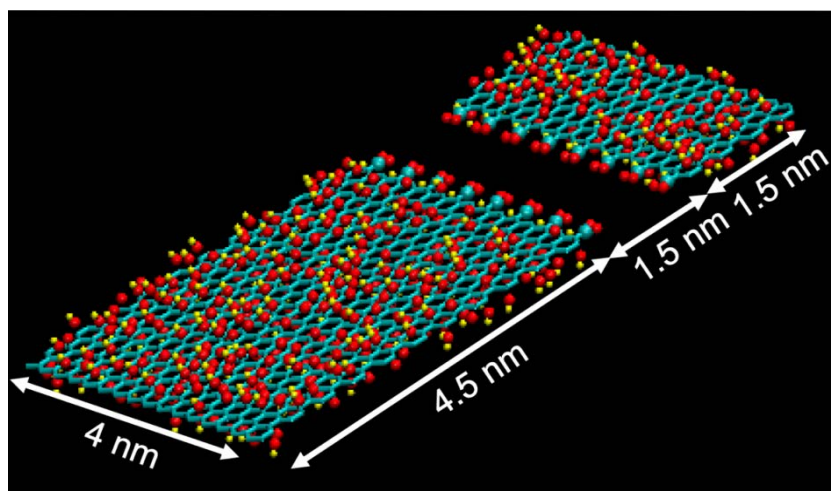

**Supplementary Fig. 15.** The Simulation model of a single slit of GO flakes. Each GO flake (7.5 ×4 nm) is generated by adding functional groups in a graphene sheet. The graphitic backbone of GO is shown in cyan lines; and carbon, oxygen and hydrogen atoms are shown in cyan, red and yellow spheres, respectively.

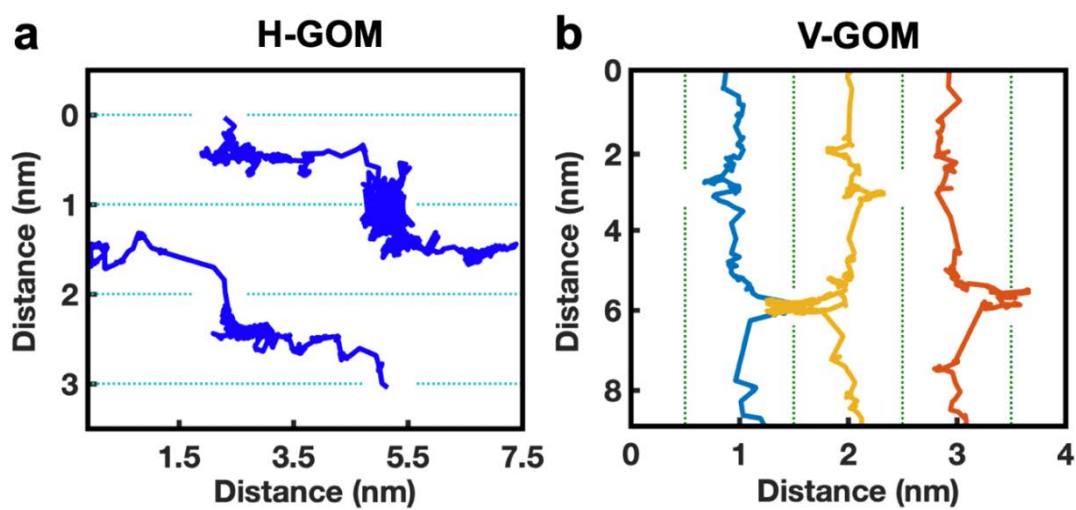

**Supplementary Fig. 16.** Representative transport trajectories of $Na^+$ ions in the H-COMs (a) and V-COMs (b). $Na^+$ ions take zigzag trajectories (solid line) to go through the gaps between adjacent GO layers (dotted lines) in H-GOMs, but pass through a single channel between adjacent GO layers (dotted lines) inside V-GOMs.

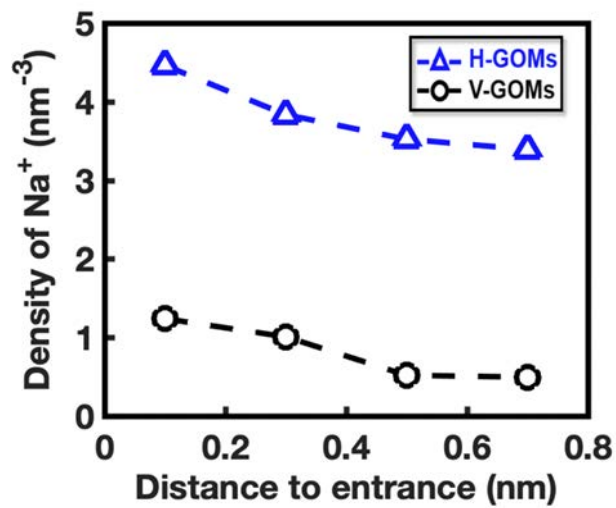

**Supplementary Fig. 17.** Comparison of the average density of Na$^+$ ions at the entrance region as a function of the distance from the entrance of H-GOMs and V-GOMs.

**Supplementary Table 1.** Energy conversions in different membranes under artificial seawater and river water conditions. *U, I* and *P<sub>max</sub>* stand for membrane potential, diffusion current and maximum output power density, respectively.

| Membrane type | Experimental condition | $U$ (mV) | $I$ (A m$^{-2}$) | $P_{max}$ (W m$^{-2}$) | Thickness (μm) | Efficiency (%) | Ref. |
|---|---|---|---|---|---|---|---|
| MKNCM | 0.5 M \| 0.01 M NaCl | - | - | 3.92 | 2 | 35 | 1 |
| UFSCNM | 0.1 M \| 0.0001 M KCl | 76 | 12 | 0.21 | 0.25 | 9.2 | 2 |
| PSS/MOF | 0.5 M \| 0.01 M NaCl | 70 | 170 | 2.87 | 1.6 | 24.4 | 3 |
| H-GOMs | 0.5 M \| 0.01 M NaCl | 160 | 20 | 0.77 | 10 | 36.6 | 4 |
| PET-BCP | 0.5 M \| 0.01 M NaCl | 75.9 | 20 | 0.35 | 13.5 | 28.5 | 5 |
| FKS | 0.5 M \| 0.01 M NaCl | 100.2 | 12.5 | 0.26 | 20 | 49.7 | 6 |
| PPy | 0.5 M \| 0.01 M NaCl | 73.4 | 4.9 | 0.087 | 20 | 26.7 | 7 |
| PCTE | 1 M \| 0.001 M KCl | 55 | 4.74 | 0.058 | 20 | 4.8 | 8 |
| IDM | 0.5 M \| 0.01 M NaCl | 73.2 | 215.7 | 3.46 | 64 | 26.5 | 6 |
| CMI | 0.5 M \| 0.01 M NaCl | 99.1 | 14.2 | 0.4 | 320 | 48.6 | 6 |
| This work | 0.5 M \| 0.01 M NaCl | 81.3 | 570 | 10.6 | 358 | 34.6 | - |

**Supplementary Text 1. Fabrication of V-GOMs.**

Graphene oxide (GO) was synthesized from purified natural graphite by the modified Hummers' method[4,9]. Graphite powder (98.5%, 8 g), sodium nitrate (NaNO$_3$, 99.0%, 8 g) and concentrated sulfuric acid (H$_2$SO$_4$, 98%, 384 ml) were completely mixed at 263.2 K for 12 hours. Then potassium permanganate (KMnO$_4$, 99.0%, 48 g) was added slowly to the mixture at 263.2 K. The mixture was firstly sufficiently stirred at 273.2 K for 1 hour, and then was reacting at 308.2 K for 12 hours. While maintaining the stirring rate and fixing temperature at 273.2 K, deionized water (1000 ml) was added slowly to the mixture (40 min) followed by the addition of hydrogen peroxide (H$_2$O$_2$, 99.0%, 10 ml, 100 min). Afterwards, the dispersion solution was treated by centrifugal process at 12,000 rpm for 20 min several times until the pH of the supernatant was approximately equal to 7. Finally, the resulting GO dispersion was freeze-dried to obtain fluffy graphene oxide (GO) and stored in dark.

GO and deionized water with a mass ratio of 1:1000 changed into dispersion by ultrasonic cell pulverizer with the power of 10 W. The dispersion was further isolated by centrifuging at 9,000 rpm for 6 min to remove the lager and insoluble GO sheets. GO dispersion (∼1 mg mL$^{-1}$) was filtrated through a cellulose ester membrane (47 mm in diameter, effective pore size of 200 nm) by the method of vacuuming[4,10]. Afterwards, the GOMs were dried in air at room temperature to remove residual water. Through this process, the horizontally stacked GOMs (H-GOMs) were obtained. To prepare V-GOMs, the H-GOMs were cut into appropriate pieces and encapsulated by epoxy glue. Exposure to ultraviolet light for 2 minutes made the epoxy glue completely cured. Through mechanical dicing, the glue formed a membrane with the thickness of about 1 mm. In addition, the membrane thickness was further reduced to 350 μm by polishing. Meanwhile, the vertically-oriented GO structures were exposed from the cured epoxy glue. Finally, the ion thinning was employed to smooth the surface of the V-GOMs.

**Supplementary Text 2. Characterization of V-GOMs.**

The effective area and thickness of the V-GOMs were characterized by scanning electron microscope (SEM). With Au deposition, the internal uniform lamellar microstructures in V-GOMs could be observed. A drop of diluted GO dispersion (0.5 µg ml$^{-1}$) was deposited on the flat surface of a mica slice and was observed by AFM (FM-Nanoview 6800AFM). The AFM observation suggests that the average thickness of GO sheets was about 0.9 nm and the lateral size distribution of these GO sheets ranged from 400 nm to 1000 nm. The lateral size distributed of the GO sheets was confirmed to be between 400 nm and 1000 nm at the concentration of 0.1 mg mL$^{-1}$ with Malvern Zetasizer NanoZS90 (Supplementary Fig. 1).

The interlayer spacing of dried GOM was tested on a polycrystalline X-ray diffractometer with a Cu Kα radiation source (Rigaku Ultima IV). X-ray diffraction (XRD) patterns showed the sharp peak at 10.3° with the scanning speed of 1° min$^{-1}$. It corresponds to the interlayer spacing of 0.86 nm, which is sufficient for sodium ions to pass.

Energy Dispersive Spectrometer (EDS) was used to analyze the types of elements in the synthesized graphene oxide. Only the elements of carbon and oxygen were observed in the EDS spectra, with the contents of 62.5% and 37.5%, respectively (Supplementary Fig. 2). It showed that the graphite was partially oxidized.

Fourier transform infrared spectroscopy (FT-IR) shows the presence of multiple oxygen-containing functional groups in graphene oxide. The characteristic GO peaks were observed at 3350, 1730, 1618, 1360, and 1050 cm$^{-1}$, which correspond to the stretching vibration of -OH, C=O, C=C, and C-O-C groups, respectively[4]. Among them, the presence of carboxyl group makes the GOMs to be cation selective.

The content of these oxygen groups can be quantitatively analyzed by X-ray photoelectron spectroscopy (XPS)[4,11]. A typical XPS pattern on GOM is shown in Supplementary Fig. 3. The total spectra (solid line) can be decomposed to four Lorentzian peaks corresponding to C=C with binding energy of 284.8 eV, C-O with binding energy of 286.7 eV, C=O with binding energy of 287.6 eV, and O=C-OH with binding energy of 289.2 eV. According to the intensity of the O=C-OH peak, the numerical density of the carboxylic acid on GO sheets can be calculated to be about 1.2% in the total carbon content. The carbon-carbon bond length in GO sheets is 0.14 nm, equivalent to 39 carbon atoms in 1 nm$^2$ area. Then, the numerical density of the carboxylic acid groups is estimated to be 0.46 nm$^{-2}$, equivalent to a surface charge density of -73.8 mC m$^{-2}$, which is in agreement with the previous work[12].

**Supplementary Text 3. Electrical Measurement.**

The membrane samples were mounted between two electrolyte cells. Each cell was filled with 10 mL NaCl solution. Varied concentrations of NaCl solution were sequentially applied at the two sides of the nanopore, constructing the concentration gradient. The IV curves of the V-GOMs system were recorded by a source meter (Keithley 6487). The degassed Milli-Q water (18.2 MΩ·cm) was used to prepare all testing solutions. To evaluate the effect of electrode potential on electrical measurement, the reference electrodes were used to apply the transmembrane potential and measure the resulting current-voltage responses of the testing V-GOMs[13]. The intercept on the vertical axis ($I_{sc}$) represents the net diffusion current. Correspondingly, the intercept on the horizontal axis ($U_{oc}$) represents the membrane potential.

**Supplementary Text 4. Energy conversion efficiency.**

Ion concentration difference is one type of Gibbs free energy that can be converted to electric power. Considering the two bulk reservoirs connected with a series of nanopores with surface charge, spontaneous ion diffusion along the concentration gradient from the high-concentration solution ($C_H$) to the low-concentration solution ($C_L$) results in the mixing of electrolyte solution. In an infinitesimal time unit ($dt$), the Gibbs free energy loss can be described as[1],

$$dG = -\frac{RT}{F} \ln \frac{a_H}{a_L} (|I_+| + |I_-|) \, dt \tag{S1}$$

where $\alpha_{H/L}$ is the chemical activity of the ionic species on $C_H$ or $C_L$ side; $I_+$ and $I_-$ is the ion flux contributed by cations and anions; F, R, T are the Faraday constant, the universal gas constant, and the temperature, respectively. In the diffusion process, the charge selectivity of nanopore causes asymmetric migration of cations and anions which can convert the Gibbs free energy partly into electric work ($dW$) in the form of net diffusion current ($I_{net}$) and transmembrane potential ($\varepsilon$),

$$dW = I_{net} \, \varepsilon \, dt \tag{S2}$$

$$I_{net} = |I_+| - |I_-| \tag{S3}$$

$$\varepsilon = (t_+ - t_-)\frac{RT}{F} \ln \frac{\alpha_H}{\alpha_L} \tag{S4}$$

The ion transference number for cations ($t_+$) and anions ($t_-$) is defined as,

$$t_+ = \frac{|I_+|}{|I_+| + |I_-|}; \quad t_- = \frac{|I_-|}{|I_+| + |I_-|} \tag{S5}$$

Then equation S2 can be simplified as,

$$dW = \frac{RT}{F} \ln \frac{a_H}{a_L} \frac{(|I_+| - |I_-|)^2}{|I_+| + |I_-|} dt \tag{S6}$$

The physical meaning of equations S1 and S6 is that, the total ion diffusion through the nanopores in an infinitesimal time unit can be characterized as the sum of the absolute value of the diffusion current contributed by cations and anions ($|I_+|+|I_-|$). Whereas, only the net diffusion current ($|I_+|-|I_-|$) generated by the asymmetric ion diffusion across the nanopore can be harvested as electric energy. When the external load equals the internal resistance of the membrane, the energy generation system achieves the maximum output power. In this case, the electric power consumed by the membrane and the external resistance is the same. Thus, the energy conversion efficiency corresponding to the maximum power generation can be calculated as:

$$\eta = \left| \frac{\frac{1}{2}dW}{dG} \right| = \frac{1}{2}\left(\frac{|I_+| - |I_-|}{|I_+| + |I_-|}\right)^2 = \frac{1}{2}(2t_+ - 1)^2 \tag{S7}$$

The calculated energy conversion efficiency is listed in Supplementary Table 1.

**Supplementary Text 5. Output power in V-GOMs.**

By mixing concentrated (500 mM) and diluted (10 mM) NaCl solution, the osmotic energy is converted to electric power through the charge separation in the ion selective nanoporous membrane. The harvested electric power can be output to an external circuit. As shown in Supplementary Fig. 4, the current densities measured in external load all decreased with the load resistance. But the output power achieves its peak value when the load resistance is equal to the internal resistance of membrane. For the V-GOMs with the varied areas of 1.37 μm × 0.98 mm, 3.11 μm × 0.96 mm and 8.82 μm × 1.02 mm, the output powers reach the maximum when the external resistance is about 100 kΩ, 47 kΩ and 15 kΩ, respectively.

**Supplementary Text 6. Energy conversion under different electrolyte conditions.**

The osmotic energy conversion in V-GOMs can be adjusted by the pH value of electrolyte solution. Alkaline solution can effectively increase power generation because the presence of $OH^-$ promotes the ionization of carboxyl group in the surface of GO sheet and accordingly enhances the ion selectivity of the V-GOMs (Supplementary Fig. 5), which consistent with previous literature reports[4]. Similarly, the suppression of ionization of carboxyl group in acidic solutions decreases the surface charge density of GO sheet. Thus, the membrane potential is the lowest in the solution with the pH value of 3. Besides the pH value, the rise of concentration difference also enhances membrane potential due to the intense diffusion under high concentration gradient.

The osmotic energy conversion in V-GOMs can be also regulated by the electrolyte ionic species. For instance, if the $Na^+$ ions are replaced by $K^+$ ions with higher mobility, the power density could be further enhanced. As shown in Supplementary Fig. 6a, the output power densities of V-GOMs reach 19.6 W m$^{-2}$ at pH=6 and 29.1 W m$^{-2}$ at pH=11 under the condition of 1000 mM|1 mM NaCl, respectively. By contrast, the output power densities in the KCl solution can achieve 23.2 W m$^{-2}$ and 34.3 W m$^{-2}$ at the same concentration and pH conditions, respectively (Supplementary Fig. 6b). The electrolyte concentration on the side of V-GOMs was fixed at 1 mM, and the concentration at the other side increased from 10 mM to 1 M.

**Supplementary Text 7. Membrane thickness dependent energy conversion.**

The tailor-made V-GOMs with different thicknesses were fixed between a pair of electrolyte solution cells. By mixing artificial seawater and river water, a source meter (Keithley 6487) was used to measure the output current. The output power ($P_R$) can be obtained by $P_R = I^2 \times R_L$. The output power density decreases linearly with the increase of membranes thickness in Supplementary Fig. 7. It shows that the energy conversion of V-GOMs has classical Ohm-like dependence. Specifically, the increment of membrane thickness linearly reduces the diffusion current; and it does not affect the membrane potential, because the change of membrane thickness does not affect the ion selectivity. The thicknesses of the tested V-GOMs were measured by SEM (Supplementary Fig. 8).

**Supplementary Text 8. Membrane area dependent energy conversion.**

By enlarging the effective membrane area, the total output power of V-GOMs can be easily promoted. What is important, the expansion of membrane area does not impair the power density. As shown in Supplementary Fig. 9, the prolonged width ($W$) raises the diffusion current linearly; and it does not affect the membrane potential, since the change in membrane area will not affect the ion selectivity. Thus, the output power can be linearly improved by enlarging the membrane area. It is beneficial for the scaling up of V-GOMs as the osmotic energy conversion systems.

The effective area of V-GOMs was characterized by SEM. To obtain the effectively average width of these V-GOMs, the statistical method was employed to characterize the width ($W$) from different V-GOMs, including $1.37 \pm 0.17$ µm (Supplementary Fig. 10), $3.11 \pm 0.19$ µm (Supplementary Fig. 11) and $8.82 \pm 0.48$ µm (Supplementary Fig. 12). Supplementary Fig. 13 shows the effective length ($L$) of different V-GOMs, including 1.01 mm, 2.11 mm and 4.46 mm. These results correspond to Fig. 2c and Fig. 3c.

**Supplementary Text 9. Similar membrane potentials between V-GOMs and H-GOMs.**

The membrane potentials in both V-GOMs and H-GOMs increase with the concentration difference (Supplementary Fig. 14). Because the charge selectivity of nanochannel is decided by the surface charge density and the geometry size, the same interlayer spacing and surface chemical groups between the V-GOMs and H-GOMs result in the similar charge selectivity. Under the same concentration difference, V-GOMs and H-GOMs show the similar membrane potential.

**Supplementary Text 10. MD simulation.**

To construct the MD simulation model of GOMs, 7.5 nm × 4 nm graphene sheets were prepared with one 1.5 nm-wide gap on the left or right edge (Supplementary Fig. 15). They were added with randomly distributed epoxy and hydroxyl functional groups on both sides by following the Lerf−Klinowski model (large fractions of hydroxyl and epoxy were bonded to carbon atoms next to each other[14,15]. All edge sites were functionalized with evenly spaced carboxylic acid groups and hydrogen atoms[16] at the liner charge density of 2 e nm$^{-1}$. One exception is that no charge was added on the outside edges of H-GOMs to prevent the ion from leaking through when applying the periodic boundary condition. The final element ratio C:H:O=52 : 16 : 32 in V (H)-GOMs and the surface charge density is 0.53 e nm$^{-2}$ in H-GOMs, which is consistent with the experimental values. Four layers of functionalized GO were stacked to form a GOM with the interlayer spacing of 1 nm. The simulation box was 4×4×20 nm$^3$ (7.6×4×9 nm$^3$) in V (H)-GOMs. The concentration of NaCl solutions on both sides of the GOM was the same as 1 M and the whole system was charge neutralized.

In the MD simulation, a 0.7 V nm$^{-1}$ voltage bias was applied along the migration direction of the ions through GOMs to accelerate simulations. Since the simulation model of GOMs is only several nanometers in size, which is much smaller than the experimental system in hundreds of micrometers, a periodic boundary condition is applied along all directions. Loading time was the time elapsed between two ions (which can go inside the GOM at least 0.8 nm) sequentially entering the GOM. The passing velocity was calculated by dividing the channel length with the average transit time.

As shown in Supplementary Fig. 16a, the ions passing through H-GOMs have to take zigzag trajectories to go through the gaps between adjacent layers. In sharp contrast, most of the ions can pass straight through a single channel inside V-GOMs (Supplementary Fig. 16b). Therefore, the ion average velocity along the permeation direction across H-GOMs is remarkably slower than that in V-GOMs. Moreover, the tortuous geometric structures in H-GOMs produce a strong barrier to impede the access of Na$^+$. It results in the evident ion enrichment at the entrance of H-GOMs (Supplementary Fig. 17).

# References

1. Zhang, Z. *et al.* Mechanically strong MXene/Kevlar nanofiber composite membranes as high-performance nanofluidic osmotic power generators. *Nat Commun* **10**, 2920 (2019).
2. Xiao, K., Giusto, P., Wen, L., Jiang, L. & Antonietti, M. Nanofluidic Ion Transport and Energy Conversion through Ultrathin Free-Standing Polymeric Carbon Nitride Membranes. *Angew. Chem. Int. Ed.* **57**, 10123-10126 (2018).
3. Li, R., Jiang, J., Liu, Q., Xie, Z. & Zhai, J. Hybrid nanochannel membrane based on polymer/MOF for high-performance salinity gradient power generation. *Nano Energy* **53**, 643-649 (2018).
4. Ji, J. et al. Osmotic Power Generation with Positively and Negatively Charged 2D Nanofluidic Membrane Pairs. *Adv. Funct. Mater.* **27**, 1603623 (2017).
5. Zhang, Z. et al. Engineered Asymmetric Heterogeneous Membrane: A Concentration-Gradient-Driven Energy Harvesting Device. *J. Am. Chem. Soc.* **137**, 14765-14772 (2015).
6. Gao, J. et al. High-performance ionic diode membrane for salinity gradient power generation. *J. Am. Chem. Soc.* **136**, 12265-12272 (2014).
7. Yu, C. et al. A smart cyto-compatible asymmetric polypyrrole membrane for salinity power generation. *Nano Energy* **53**, 475-482 (2018).
8. Kwon, K., Lee, S. J., Li, L., Han, C. & Kim, D. Energy harvesting system using reverse electrodialysis with nanoporous polycarbonate track-etch membranes. Int. J. Energy Res. *Int. J. Energy Res.* **38**, 530-537 (2014).
9. Huang, H. et al. Ultrafast viscous water flow through nanostrand-channelled graphene oxide membranes. *Nat. Commun.* **4**, 2979 (2013).
10. Dikin, D. A. et al. Preparation and characterization of graphene oxide paper. *Nature* **448**, 457-460 (2007).
11. Chen, L. et al. Ion sieving in graphene oxide membranes via cationic control of interlayer spacing. *Nature* **550**, 1-4 (2017).
12. Kotov, N. A., Dekany, I. & Fendler, J. H. Ultrathin Graphite Oxide-Polyelectrolyte Composites Prepared by Self-Assembly: Transition Between Conductive and Non-Conductive States. *Adv. Mater.* **8**, 637-641 (1996).
13. Guo, W. et al. Energy Harvesting with Single-Ion-Selective Nanopores: A Concentration-Gradient-Driven Nanofluidic Power Source. *Adv. Funct. Mater.* **20**, 1339-1344 (2010).
14. Lerf, A., He, H., Forster, M. & Klinowski, J. Structure of Graphite Oxide Revisited. *J. Phys. Chem. B* **102**, 4477-4482 (1998).
15. Dai, H., Xu, Z. & Yang, X. Water Permeation and Ion Rejection in Layer-by-Layer Stacked Graphene Oxide Nanochannels: A Molecular Dynamics Simulation. *J. Phys. Chem. C* **120**, 22585-22596 (2016).
16. Williams, C. D., Carbone, P. & Siperstein, F. R. Computational characterisation of dried and hydrated graphene oxide membranes. *Nanoscale* **10**, 1946-1956 (2018).